\documentclass[twocolumn,showpacs,preprintnumbers,amsmath,aps]{revtex4}
\usepackage{graphicx}
\usepackage{dcolumn}
\usepackage{bm}
\begin{document}
\newcommand{\kvec}{\mbox{{\scriptsize {\bf k}}}}
\def\eq#1{(\ref{#1})}
\def\fig#1{Fig.\hspace{1mm}\ref{#1}}
\def\tab#1{table\hspace{1mm}\ref{#1}}
\title{
---------------------------------------------------------------------------------------------------------------\\
Properties of the superconducting state in molecular metallic hydrogen under pressure at 347 GPa}
\author{R. Szcz{\c{e}}{\`s}niak, M.W. Jarosik}
\affiliation{Institute of Physics, Cz{\c{e}}stochowa University of Technology, Al. Armii Krajowej 19, 42-200 Cz{\c{e}}stochowa, Poland}
\email{jarosikmw@wip.pcz.pl}
\date{\today} 
\begin{abstract}
The thermodynamic properties of the superconducting state induced in metallic molecular hydrogen under the influence of pressure $347$ GPa were determined. In particular, it has been shown that the critical temperature ($T_{C}$) changes in the range from $120$ K to $90$ K for $\mu^{*}\in\left<0.08,0.15\right>$, where $\mu^{*}$ is the value of the Coulomb pseudopotential. Next, the energy gap near the temperature
of zero Kelvin ($2\Delta\left(0\right)$) was calculated. It has been stated, that the dimensionless ratio $2\Delta\left(0\right)/k_{B}T_{C}$ slightly decreases with the increase of $\mu^{*}$ from $3.98$ to $3.84$. In the last step, the ratio of effective electron mass ($m^{*}_{e}$) to the bare electron  mass ($m_{e}$)) was determined. It has been proved that $m^{*}_{e}/m_{e}$ takes its highest value equal to $1.96$ for $T=T_{C}$.
\end{abstract}
\pacs{74.20.Fg, 74.25.Bt, 74.62.Fj}
\maketitle
%
\section{Introduction}

The influence of the pressure ($p$) on thermodynamic properties of the superconducting state is one of more interesting issues in the solid state physics.

In the case of simple-metal superconductors with the electron-phonon pairing mechanism for a long time there was conviction, that the increase of the pressure leads to the decrease of the critical temperature ($T_{C}$) and the remaining thermodynamic parameters do not demonstrate interesting properties \cite{Schilling}. The change took place in the last years when in some of the classical materials the pressure-induced superconducting state of relatively high critical temperature was discovered, for example: in lithium $\left[T_{C}\right]_{\rm max}\simeq 14$ K for $p\simeq 30$ GPa and in calcium, where $\left[T_{C}\right]_{\rm max}\simeq 25$ K for $p\simeq 160$ GPa \cite{lit}, \cite{wapn}.

The obtained results are keeping an eye on the metallic hydrogen again, where the existence of the pressure-induced superconducting
state with the very high value of the critical temperature (comparable with the value of the room temperature) was expected \cite{Ashcroft}. 
From the theoretical point of view the following facts enable to make the above assumption true \cite{Maksimov}: (i) high value of the phonon frequency caused by a small mass of atomic nuclei building the crystal lattice (single proton), (ii) large value of the electron-phonon coupling related to the lack of the inner electronic shells and (iii) low value of the Coulomb pseudopotential (at least in the range of pressures up to $500$ GPa  \cite{Richardson}, \cite{Szczesniak2}).

Since the time of publishing the Ashcroft's work it is stated that the value of $T_{C}$ for metallic hydrogen is located in the range from $80$ K to $300$ K for selected pressures lesser than $500$ GPa \cite{Richardson}, \cite{Caron}, \cite{Zhang}. Under the extremely high pressure ($p=2000$ GPa) the metallic state of the hydrogen may transit itself into the superconducting state in the range of the temperatures from $413$ K to $631$ K \cite{Maksimov}, \cite{Szczesniak2}. Let us note that in this case the thermodynamic properties of the superconducting state strongly deviate from the predictions of the BCS theory \cite{BCS}.

In the range of the "low" pressures (up to $500$ GPa) the thermodynamic properties outside the critical temperature were not studied. For that reason we will calculate the selected thermodynamic parameters of the superconducting state in the molecular metallic hydrogen for the exemplary pressure of $347$ GPa. In particular, in the framework of the Eliashberg approach \cite{Eliashberg}, we will determine the value of the critical temperature as a function
of the Coulomb pseudopotential. Next, we will exactly calculate the order parameter ($\Delta\left(T\right)$) and the wave function renormalization factor. On the basis of obtained results the ratio $2\Delta\left(0\right)/k_{B}T_{C}$ as a function of the Coulomb pseudopotential and the maximum value of the electron effective mass will be determined. Additionally, in the paper, the relationship between the structure of the electron-phonon coupling and the form of the solutions to the Eliashberg equations along the real axis will be discussed.
 
\section{The Eliashberg equations}

The Eliashberg equations can be determined on the real axis, imaginary axis or in the mixed representation (simultaneously on the real and imaginary axis). Each of the mentioned approaches have the unique advantages, but also come with the characteristic mathematical problems. The exact solution of the Eliashberg equations on the real axis composes an extremely difficult mathematical problem \cite{Zheng}. From the other side, the Eliahberg equations on the imaginary axis can be solved in a simpler way \cite{Szczesniak3}. However, in this case in order to interpret physically the obtained results (e.g. to calculate the value of the energy gap near the temperature of zero Kelvin) the achieved solutions should be analytically continued on the real axis. Unfortunately, the procedure of the analytical continuation is complicated and it demands very high precision during the numerical calculations \cite{AnalitycznaKontynuacja}. It is worth mentioning, that the results obtained in a such way are stable only in the range of the low frequencies.

The Eliashberg equations in the mixed representation are proclaimed to be a reasonable compromise between the two previously mentioned approaches. Firstly, they can be exactly solved in the much simpler way than equations on the real axis. Secondly, the obtained results are stable even for the very large values of the frequency. In the case of molecular metallic hydrogen it is an important matter, because the maximum phonon frequency  ($\Omega_{{\rm max}}$) is equal to $477$ meV \cite{Zhang}.    

The Eliashberg equations in the mixed representation can be written in the form \cite{Eliashberg}, \cite{Marsiglio}:
%
\begin{widetext}
\begin{eqnarray}
\label{r1}
\phi\left(\omega+i\delta\right)&=&
                                  \frac{\pi}{\beta}\sum_{m=-M}^{M}
                                  \left[\lambda\left(\omega-i\omega_{m}\right)-\mu^{*}\theta\left(\omega_{c}-|\omega_{m}|\right)\right]
                                  \frac{\phi\left(i\omega_{m}\right)}
                                  {\sqrt{\omega_m^2Z^{2}\left(i\omega_{m}\right)+\phi^{2}\left(i\omega_{m}\right)}}\\ \nonumber
                              &+& i\pi\int_{0}^{+\infty}d\omega^{'}\alpha^{2}F\left(\omega^{'}\right)
                                  \left[\left[N\left(\omega^{'}\right)+f\left(\omega^{'}-\omega\right)\right]
                                  \frac{\phi\left(\omega-\omega^{'}+i\delta\right)}
                                  {\sqrt{\left(\omega-\omega^{'}\right)^{2}Z^{2}\left(\omega-\omega^{'}+i\delta\right)
                                  -\phi^{2}\left(\omega-\omega^{'}+i\delta\right)}}\right]\\ \nonumber
                              &+& i\pi\int_{0}^{+\infty}d\omega^{'}\alpha^{2}F\left(\omega^{'}\right)
                                  \left[\left[N\left(\omega^{'}\right)+f\left(\omega^{'}+\omega\right)\right]
                                  \frac{\phi\left(\omega+\omega^{'}+i\delta\right)}
                                  {\sqrt{\left(\omega+\omega^{'}\right)^{2}Z^{2}\left(\omega+\omega^{'}+i\delta\right)
                                  -\phi^{2}\left(\omega+\omega^{'}+i\delta\right)}}\right]
\end{eqnarray}
and
\begin{eqnarray}
\label{r2}
Z\left(\omega+i\delta\right)&=&
                                  1+\frac{i\pi}{\omega\beta}\sum_{m=-M}^{M}
                                  \lambda\left(\omega-i\omega_{m}\right)
                                  \frac{\omega_{m}Z\left(i\omega_{m}\right)}
                                  {\sqrt{\omega_m^2Z^{2}\left(i\omega_{m}\right)+\phi^{2}\left(i\omega_{m}\right)}}\\ \nonumber
                              &+&\frac{i\pi}{\omega}\int_{0}^{+\infty}d\omega^{'}\alpha^{2}F\left(\omega^{'}\right)
                                  \left[\left[N\left(\omega^{'}\right)+f\left(\omega^{'}-\omega\right)\right]
                                  \frac{\left(\omega-\omega^{'}\right)Z\left(\omega-\omega^{'}+i\delta\right)}
                                  {\sqrt{\left(\omega-\omega^{'}\right)^{2}Z^{2}\left(\omega-\omega^{'}+i\delta\right)
                                  -\phi^{2}\left(\omega-\omega^{'}+i\delta\right)}}\right]\\ \nonumber
                              &+&\frac{i\pi}{\omega}\int_{0}^{+\infty}d\omega^{'}\alpha^{2}F\left(\omega^{'}\right)
                                  \left[\left[N\left(\omega^{'}\right)+f\left(\omega^{'}+\omega\right)\right]
                                  \frac{\left(\omega+\omega^{'}\right)Z\left(\omega+\omega^{'}+i\delta\right)}
                                  {\sqrt{\left(\omega+\omega^{'}\right)^{2}Z^{2}\left(\omega+\omega^{'}+i\delta\right)
                                  -\phi^{2}\left(\omega+\omega^{'}+i\delta\right)}}\right]. 
\end{eqnarray}
\end{widetext}
%

In Eqs. \eq{r1} and \eq{r2} the symbols $\phi$ and $Z$ denote the order parameter function and the wave function renormalization factor determined respectively on the real axis ($\omega$) or imaginary axis ($\omega_{m}$), where $\omega_{m}\equiv \left(\pi / \beta\right)\left(2m-1\right)$ is the $m$-th Matsubara frequency and $\beta\equiv\left(k_{B}T\right)^{-1}$ ($k_{B}$ is the Boltzmann constant). The order parameter on the real axis ($\Delta\left(\omega\right)$) is defined in a following way: $\Delta\left(\omega\right)\equiv\phi\left(\omega\right)/Z\left(\omega\right)$. The pairing kernel of the electron-phonon interaction is given by the expression:
\begin{equation}
\label{r3}
\lambda\left(z\right)\equiv 2\int_0^{\Omega_{\rm{max}}}d\Omega\frac{\alpha^{2}F\left(\Omega\right)\Omega}{\Omega^2-z^2},
\end{equation}
where the Eliashberg function ($\alpha^{2}F\left(\Omega\right)$) for the molecular metallic hydrogen at the pressure of $347$ GPa was determined in the paper \cite{Zhang}. Symbol $\mu^{*}$ denotes the value of the Coulomb pseudopotential; $\Theta$ is the Heaviside unit function and $\omega_{c}$ is called the phonon cut-off frequency ($\omega_{c}=3\Omega_{\rm{max}}$). In the presented work, $\mu^{*}$ was taken into account parametrically. In particular, it was assumed that: $\mu^{*}\in\left<0.08,0.15\right>$. In the last two terms of Eqs. \eq{r1} and \eq{r2} the symbols $N\left(\omega\right)$ and $f\left(\omega\right)$ denote the statistical functions of bosons and fermions respectively.

The Eliashberg equations were solved numerically for $M=800$. In this case the stability of all solutions was assured for the minimal temperature equal to $23.2$ K ($2$ meV) regardless of the assumed value of $\mu^{*}$. 

\section{The numerical and analytical results}
%
\subsection{Value of the critical temperature}
%
\begin{figure}[t]%
\includegraphics*[scale=0.31]{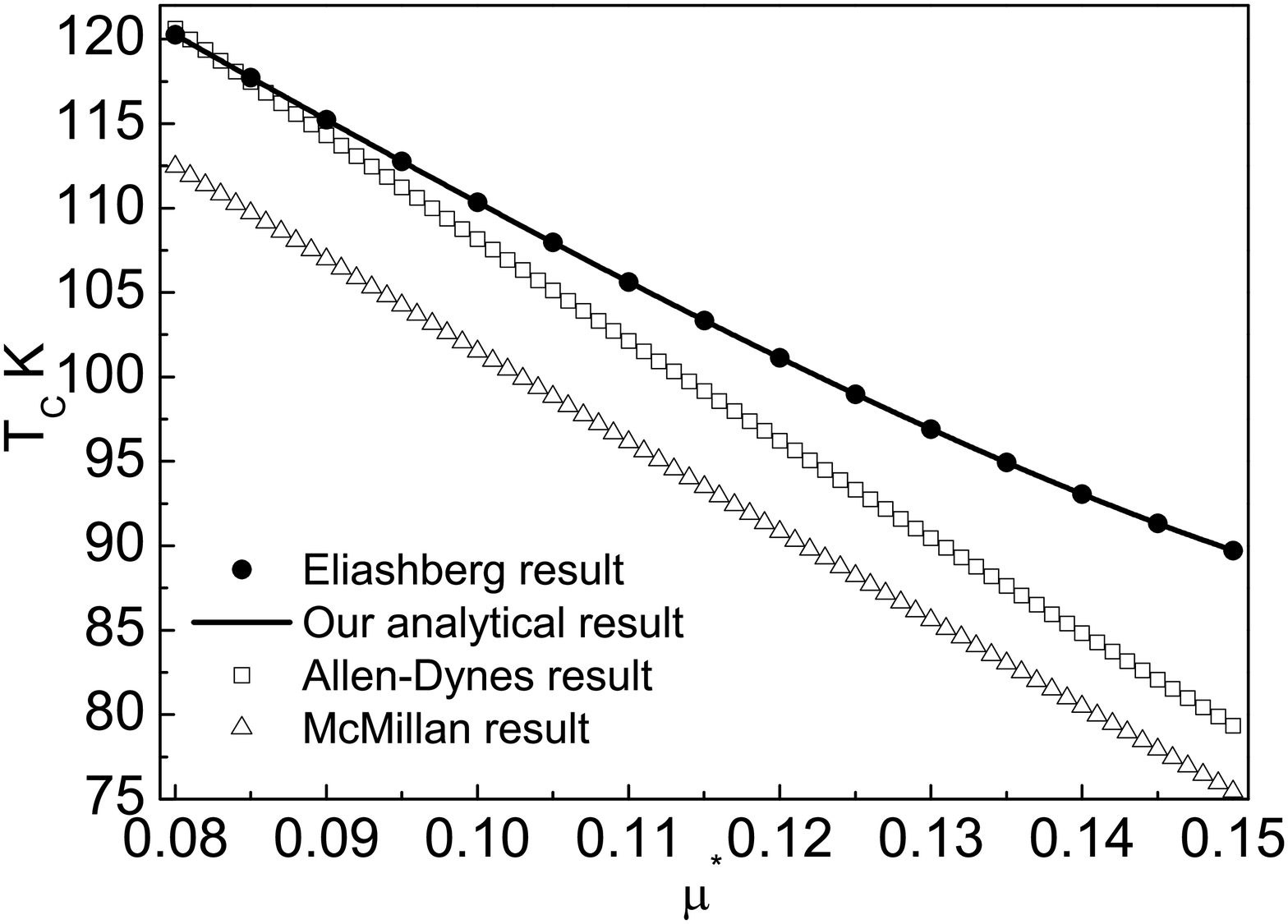}
\caption{The value of the critical temperature as a function of the Coulomb pseudopotential. The filled circles represent the exact results obtained with an use of the Eliashberg equations. The solid line has been obtained using the {\it modified} Allen-Dynes formula. Empty squares and triangles refer respectively to the classical Allen-Dynes expression and the McMillan formula ($f_{1}=f_{2}=1$) \cite{AllenDynes}, \cite{McMillan}.}
\label{f1}
\end{figure}
%

The dependence of the critical temperature on the Coulomb pseudopotential for the low-$T_{C}$ superconductors can be determined in the easiest way when using the formula drawn out by Allen and Dynes \cite{AllenDynes}. Unfortunately, in the case of the metallic hydrogen, the classical formula determines correctly $T_{C}\left(\mu^{*}\right)$ only for the very low values of $\mu^{*}$ (see \fig{f1}); for the highest values of the pseudopotential the critical temperature is considerably underestimated. For that reason selected parameters that appear in a classical expression were calculated once again. In particular, we have used the least-squares analysis and $120$ values of $T_{C}\left(\mu^{*}\right)$ obtained with a help of the Eliashberg equations. In the considered case, the {\it modified} Allen-Dynes formula takes the form:
\begin{equation}
\label{r4}
k_{B}T_{C}=f_{1}f_{2}\frac{\omega_{\rm ln}}{1.2}\exp\left[\frac{-1.04\left(1+\lambda\right)}{\lambda-\mu^{*}\left(1+0.62\lambda\right)}\right],
\end{equation}
where the strong-coupling correction function ($f_{1}$) and the shape correction function ($f_{2}$) are given by:
\begin{equation}
\label{r5}
f_{1}\equiv\left[1+\left(\frac{\lambda}{\Lambda_{1}}\right)^{\frac{3}{2}}\right]^{\frac{1}{3}}
\end{equation}
and
\begin{equation}
\label{r6}
f_{2}\equiv 1+\frac
{\left(\frac{\sqrt{\omega_{2}}}{\omega_{\rm{ln}}}-1\right)\lambda^{2}}
{\lambda^{2}+\Lambda^{2}_{2}}.
\end{equation}
The new functions $\Lambda_{1}$ and $\Lambda_{2}$ can be written as follows: 
\begin{equation}
\label{r7}
\Lambda_{1}\equiv 4.11\left(1-4.61\mu^{*}\right)
\end{equation}
and 
\begin{equation}
\label{r8}
\Lambda_{2}\equiv 13.57\left(1-4.79\mu^{*}\right)\left(\frac{\sqrt{\omega_{2}}}{\omega_{\rm{ln}}}\right).
\end{equation}
The parameter $\omega_{2}$ denotes the second moment of the normalized weight function and should be calculated on the basis of the formula:
\begin{equation}
\label{r9}
\omega_{2}\equiv 
\frac{2}{\lambda}
\int^{\Omega_{\rm{max}}}_{0}d\Omega\alpha^{2}F\left(\Omega\right)\Omega.
\end{equation}
The quantity $\omega_{{\rm ln}}$ is called the logarithmic phonon frequency and $\lambda$ is the electron-phonon coupling constant:
\begin{equation}
\label{r10}
\omega_{{\rm ln}}\equiv \exp\left[\frac{2}{\lambda}
\int^{\Omega_{\rm{max}}}_{0}d\Omega\frac{\alpha^{2}F\left(\Omega\right)}
{\Omega}\ln\left(\Omega\right)\right],
\end{equation}
\begin{equation}
\label{r11}
\lambda\equiv 2\int^{\Omega_{\rm{max}}}_{0}d\Omega\frac{\alpha^{2}F\left(\Omega\right)}{\Omega}.
\end{equation}
In the case of molecular metallic hydrogen ($p=347$ GPa) the following results were achieved: $\sqrt{\omega_{2}}=199$ $\rm{meV}$, $\omega_{{\rm ln}}=142$ meV and $\lambda=0.927$.

Below, we draw the reader's attention to the fact, that the new parameterization of $\Lambda_{1}$ and $\Lambda_{2}$ very relevantly affects the dependence of the functions $f_{1}$ and $f_{2}$ on the value of the Coulomb pseudopotential. Namely, it comes to: $f_{1}>f_{2}\simeq 1$. Additionally, the values of $f_{1}$ grow together with the growth of the pseudopotential. The above result means that $T_{C}$ very significantly depends on the strong-coupling effects modeled by the function $f_{1}$; while the shape correction function weakly affects the critical temperature.
          
When coming back to the data presented in \fig{f1} it can be easily noticed, that in the considered range of the Coulomb pseudopotential's values the critical temperature decreases from $120$ K to $90$ K. The achieved result clearly states that, even for the relatively large $\mu^{*}$, the critical temperature takes very high value.

\subsection{The order parameter function}

\begin{figure}[t]%
\includegraphics*[scale=0.31]{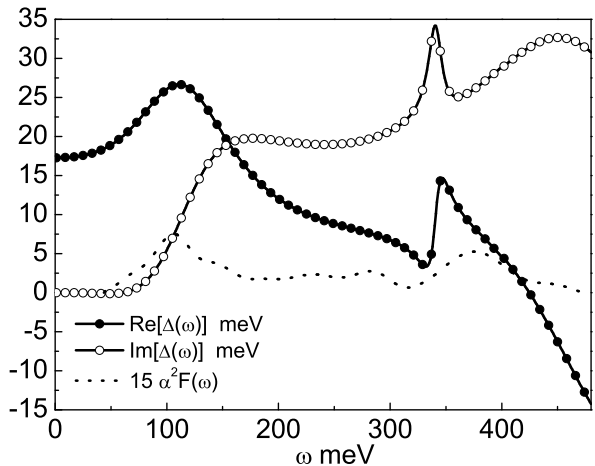}
\caption{The real and imaginary part of the order parameter as a function of the frequency for $T=58$ K (5 meV) and $\mu^{*}=0.1$. In the figure there is also plotted the rescaled Eliashberg function.}
\label{f2}
\end{figure}

The Eliashberg equations were solved for the range of temperature from $23.2$ K to $T_{C}$. As an example, the dependence of the real and imaginary part of the order parameter on the frequency in \fig{f2} was presented. For the low values of $\omega$ the non-zero is only the real part of the order parameter. Next, we observe the characteristic sequence of the local maximums (minimums) and the points of the fold both for Re[$\Delta\left(\omega\right)$] and Im[$\Delta\left(\omega\right)$]. Additionally, in \fig{f2} the Eliashberg function is shown; in order of the better representation its values were multiplied by $15$. When comparing the shapes of Re[$\Delta\left(\omega\right)$] and Im[$\Delta\left(\omega\right)$] with the form of the Eliashberg function the correlation between the characteristic points of the considered functions can be easily noticed. Let us mark the fact, that from the physical point of view, the existence of the peak in the Eliashberg function determines the area of $\omega$ in which the electron-phonon interaction is exceptionally strong. Thus, the distribution of the characteristic points of the real and imaginary part of the order parameter is inseparably connected with the specific structure of the electron-phonon interaction.
%
%

\begin{figure}[t]%
\includegraphics*[scale=0.31]{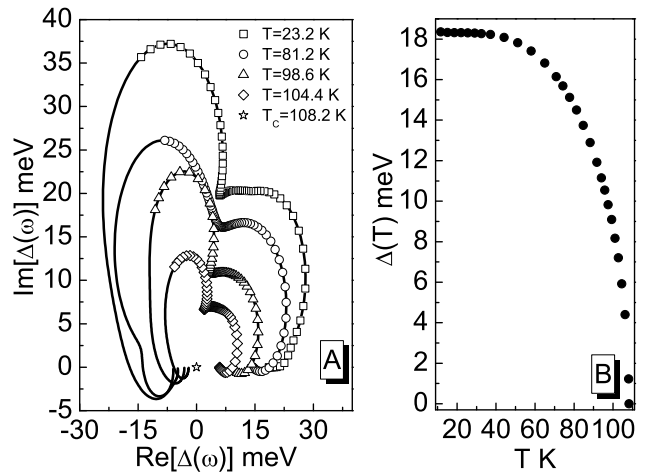}
\caption{(A) The order parameter on the complex plane for the selected values of the temperature. The solid lines with points were obtained for $\omega\in\left<0, \Omega_{\rm max}\right>$; the solid lines represent the results for the frequency from $\Omega_{\rm max}$ to $1600$ meV. (B) The dependence of the order parameter on the temperature. In both cases $\mu^{*}=0.1$ was assumed.}
\label{f3}
\end{figure}

In a global way the dependence of $\Delta\left(\omega\right)$ on the temperature was presented in \fig{f3} (A). It can be clearly seen that on the complex plane the values of $\Delta\left(\omega\right)$ form the characteristic "ear". The obtained result allows to characterize the effective potential of the electron-electron interaction, which is connected with the real part of the order parameter \cite{Varelogiannis}. In particular, let us notice that only in the range of frequencies from zero to the frequency slightly lesser than $\Omega_{\rm max}$, the effective electron-electron interaction is attractive (Re$\left[\Delta\left(\omega\right)\right]>0$); the achieved conclusion is true for any considered value of $\mu^{*}$.

Below, the values of the ratio $R_{1}\equiv 2\Delta\left(0\right)/k_{B}T_{C}$ for $\mu^{*}\in\left<0.08,0.15\right>$ are determined; the symbol $\Delta\left(0\right)$ denotes the order parameter near the temperature of zero Kelvin. Let us notice, that in the framework of the weak-coupling approach (the BCS model) the parameter $R_{1}$ takes the constant value equal to $3.53$. In the case when $\lambda>0.2$ its exact value can be calculated only with the help of the Eliashberg equations. In particular, $\Delta\left(T\right)$ is calculated on the basis of  Re$\left[\Delta\left(\omega\right)\right]$ with an use of the expression: 
\begin{equation}
\label{r12}
\Delta\left(T\right)={\rm Re}\left[\Delta\left(\omega=\Delta\left(T\right)\right)\right].
\end{equation}

\begin{figure}[t]%
\includegraphics*[scale=0.31]{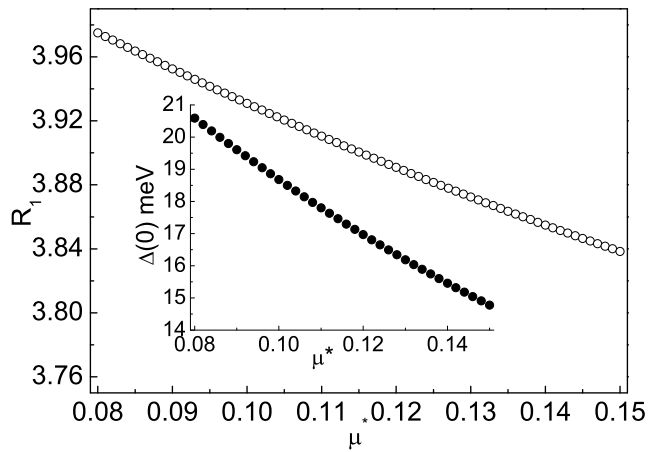}
\caption{Dimensionless ratio $R_{1}$ as a function of $\mu^{*}$. The dependence of the parameter $\Delta\left(0\right)$ on the Coulomb pseudopotential has been shown in the figure's inset.}
\label{f4}
\end{figure}

In \fig{f3} (B) the dependence of the order parameter on $T$ is plotted. Due to the fast saturation of the function $\Delta\left(T\right)$ in the area of the lower temperatures, for $\Delta\left(0\right)$ one can assume the value of $\Delta\left(T=23.2 {\rm K}\right)$. The final results are presented in \fig{f4}, where the function $R_{1}\left(\mu^{*}\right)$ is shown. It is easy to notice, that the values of the ratio $R_{1}$ are higher than $\left[R_{1}\right]_{{\rm BCS}}$ and only slightly decreasing with the growth of the Coulomb pseudopotential (from $3.98$ to $3.84$). The determined dependency of $R_{1}$ on $\mu^{*}$ is connected with the strong, but comparable, influence of the electronic depairing correlations on $T_{C}$ and $\Delta\left(0\right)$; see \fig{f1} and the inset in \fig{f4}.     
%
%
%

\subsection{The electron effective mass}

\begin{figure}[t]%
\includegraphics*[scale=0.31]{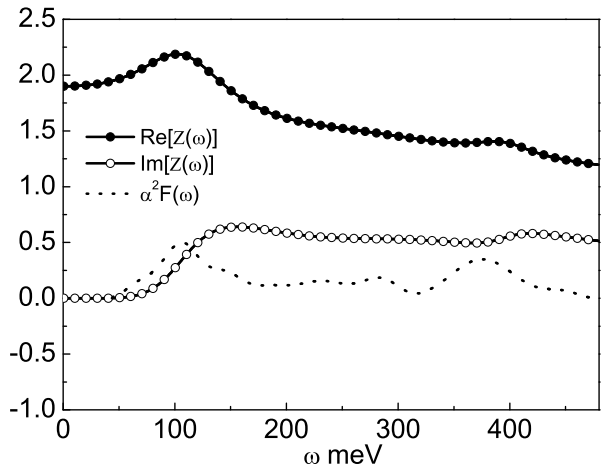}
\caption{The real and imaginary part of the wave function renormalization factor as a function of frequency for $T=58$ K and  $\mu^{*}=0.1$. 
In the figure the Eliashberg function is also plotted.}
\label{f5}
\end{figure}

\begin{figure}[t]%
\includegraphics*[scale=0.31]{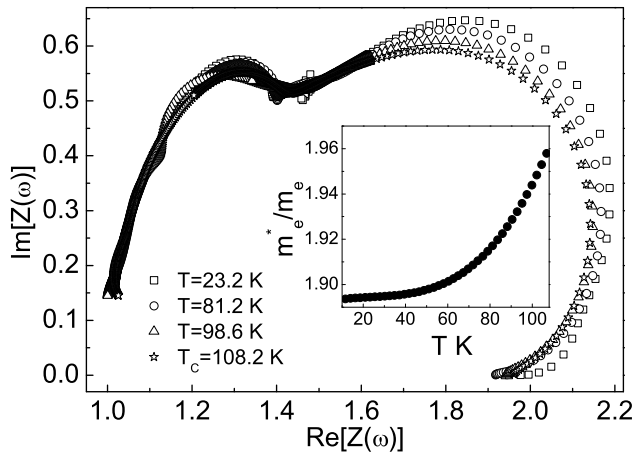}
\caption{The wave function renormalization factor on the complex plane for the selected values of $T$. In the inset the dependence of $m^{*}_{e}/m_{e}$ on the temperature has been marked. In both cases $\mu^{*}=0.1$ was assumed.}
\label{f6}
\end{figure}

In \fig{f5} the wave function renormalization factor is plotted. Similarly as for the order parameter, in the low-frequencies region the non-zero is only the real part of $Z\left(\omega\right)$. For higher frequencies, we can observe the complicated dependence of Re$\left[Z\right]$ and Im$\left[Z\right]$ on $\omega$, which is plainly correlated with the form of the Eliashberg function. However, it should be noticed, that in the opposition to the order parameter, the function $Z\left(\omega\right)$ is significantly weaker dependent on the temperature (see \fig{f6}).
%
%
%

In the framework of the Eliashberg theory the wave function renormalization factor plays a very significant role, because with the help of it one can calculate the ratio of the electron effective mass ($m^{*}_{e}$) to the bare electron mass ($m_{e}$):
\begin{equation}
\label{r13}
\frac{m^{*}_{e}}{m_{e}}={\rm Re}\left[Z\left(0\right)\right].
\end{equation}

In order to precisely track the values of the electron effective mass in the superconducting area the dependence of  $m^{*}_{e}/m_{e}$ on the temperature is plotted in the inset of \fig{f6}. For the molecular metallic hydrogen the value of the electron effective mass is pretty high and takes its maximum equal to $1.96$ for $T=T_{C}$. Let us notice that in this case $m^{*}_{e}$ is independent of $\mu^{*}$. 
 
\section{Summary}

In this paper the thermodynamic properties of the superconducting state induced in molecular metallic hydrogen ($p=347$ GPa) were studied. In order to do that, the Eliashberg equations in the mixed representation were exactly solved. The obtained results enabled to determine the dependence of the order parameter and the wave function renormalization factor in the wide range of the frequency. It has been stated, that for $\omega\in\left<0,\Omega_{\rm max}\right>$ the existence of the characteristic points of the function $\Delta\left(\omega\right)$ and $Z\left(\omega\right)$ is strictly correlated with the structure of the electron-phonon coupling modeled by the Eliashberg function. Additionally, in the range of frequencies from zero to the frequency slightly lesser than $\Omega_{\rm max}$, the effective electron-electron interaction is attractive (Re$\left[\Delta\left(\omega\right)\right]>0$).

When basing on the exact solutions of the Eliashberg equations it has been shown, that the value of the critical temperature decreases slower with the growth of the Coulomb pseudopotential, than it is predicted by the classical Allen-Dynes formula. In particular, $T_{C}$ changes from $120$ K to $90$ K for $\mu^{*}\in\left<0.08,015\right>$. Next, the dependence of the dimensionless ratio $R_{1}$ on $\mu^{*}$ was determined. It has been proven that the parameter $R_{1}$ weakly depends on the value of the Coulomb pseudopotential and $\left[R_{1}\right]_{{\rm max}}=3.98$ for $\mu^{*}=0.08$. In the last step, the value of the electron effective mass in the superconducting area was calculated. On the base of the obtained results it has been shown, that $\left[m^{*}_{e}/m_{e}\right]_{\rm max}=1.96$ for $T=T_{C}$.

\begin{acknowledgments}
The authors wish to thank Prof. K. Dzili{\'n}ski for providing excellent working conditions and the financial support; Mr A.P. Durajski for the productive scientific discussion that improved the quality of the presented work. All numerical calculations were based on the Eliashberg function sent to us by: {\bf L. Zhang}, Y. Niu, Q. Li, T. Cui, Y. Wang, {\bf Y. Ma}, Z. He and G. Zou for whom we are also very thankful. 
\end{acknowledgments}
%

%
\end{document}